\documentstyle[iopconf1,epsf]{article}
\begin{document}
\title{Dark Halos and Galaxy Evolution}

\author{Claudio Firmani\footnote{Email: firmani@astroscu.unam.mx} and 
Vladimir Avila-Reese\footnote{Email: avila@astroscu.unam.mx}}

\affil{Instituto de Astronom\'{\i}a-UNAM, A.P. 70-264, 04510 M\'{e}xico
D.F., and Dept. of Astronomy-NMSU, P.O. Box 30001, Las Cruces, NM 88003-8001}

\beginabstract
We study the evolution of disk galaxies within the frame of the
cold dark matter (CDM) cosmologies. The hydrodynamics of a centrifugally
supported gaseous disk and the growth of a stellar disk is calculated in
detail taking into account the energy balance of the ISM and the gravitational 
instabilities that concern gas and stars. The halo density profile is
derived from the primordial cosmological conditions and its
gravitational contraction produced by the disk is included. Several
features of the spiral galaxies at different
redshifts are predicted, and the main factors which influence on these
features are found. A strong evidence is provided that the Tully-Fisher
(TF) relation is an imprint of the primordial cosmological conditions.
\endabstract

\section{Introduction}
Individual galaxies provide useful information on the physical properties of
dark matter (DM).  However, it is necessary first to
clarify the link between the luminous matter of the galaxies and their 
DM halos. In the
general problem of how galaxies do form and evolve in a cosmological
frame little progress has been done regarding the internal physics of 
galaxies. We remark three categories of theoretical approaches to the
problem of galaxy formation and evolution. 1) \textit{The analytical
models} (Mo, Mao \& White 1998; see also Dalcanton, Spergel, \& Summer
1997). At a given time an exponential thin disk with a specific angular
momentum is introduced in a DM halo taking into account the gravitational
contraction produced by the disk on the halo. The halo properties are
taken from the results of N-body simulations. This approach is the most
economical and is particularly useful and elegant to study the properties of
galaxy populations at different redshifts, as well as some global
properties of disk galaxies as the baryon fraction involved in the
disk and its angular momentum. However with this approach it was not
possible to follow the individual evolution of galaxies. 2)
\textit{The semi-analytical models} (Kauffmann, White \& Guiderdoni
1993; Cole \etal 1994; Somerville \& Primack 1998, and more references 
therein). In this approach the metabolism of a galaxy is basically
identified in the disk-halo feedback. The luminous galaxy is modeled
through recipes characterized by free parameters. This method allows
to predict several global properties for a given galaxy
as well as for a population of galaxies on the basis of their merging 
histories,
but although simplified galaxy structural properties may be
introduced, it does not allow to predict the internal properties of
galaxies. 3) \textit{The numerical simulations} (e.g., Yepes 1997, and 
the references therein). This method is very predictive to identify halo
properties and is very promissory at future, but at galactic scales technical
difficulties related to the resolution and the processes that involve gas
and stars make the results of this method currently the less predictive.
Because of the reasons listed before we address the problem of galaxy
evolution in a cosmological frame, starting with a paradigm which takes into
account the internal physics of a galaxy.

\section{The method}
The kernel of this work is a \textit{semi-numerical} approach where several
internal processes of the disk, not considered in the other methods
mentioned before, are taken into account. We start modeling the
physics of a galactic disk where the hydrodynamics of the gas and the
stars is taken into account with axial symmetry. A secular bulge
formation is introduced applying a local gravitational
stability criterion to the stellar disk. The star formation (SF) is
induced in the disk
by gravitational instabilities and it is self-regulated by an
energetic balance in the ISM where the main energy source is due to
SNs; simple stellar population synthesis models are considered
(Firmani \& Tutukov 1994; Firmani, Hern\'andez \& Gallagher
1996). The galactic evolutionary models are inserted in a cosmological 
background: the structure and evolution of the DM halos which surround
the disks and the mass accretion rates over them are calculated from
initial conditions defined by the cosmological model, which is
specified by the mass fractions of the DM species (cold $\Omega _{CDM}$, 
hot $\Omega _{\nu}$), the vacuum energy ($\Omega _\Lambda $), 
and the baryon matter ($\Omega _b$), and by the value of the Hubble
constant ($h=H_0/100Kms^{-1}Mpc^{-1}$). We have taken as the
representative models the $SCDM$ ($\Omega _{CDM}=0.96$,
$h=0.5$), the $\Lambda CDM$ ($\Omega _{CDM}=0.327$, 
$\Omega _\Lambda =0.65$, $h=0.65$), the $OCDM$ ($\Omega _{CDM}=0.327$,
$h=0.65$),and the $H+CDM$ ($\Omega _{CDM}=0.94$, $\Omega _{\nu}=0.2$, 
$\Omega _b=0.06$, $h=0.5$). The value of $\Omega _b$ was taken equal
to $0.01h^2$ where it is not specified. The primordial density fluctuation
field is characterized by a gaussian statistical distribution with a power
spectrum taken from Sugiyama (1996) and normalized to the COBE
data. 
\begin{figure}
\vspace*{6.7cm}
\includegraphics{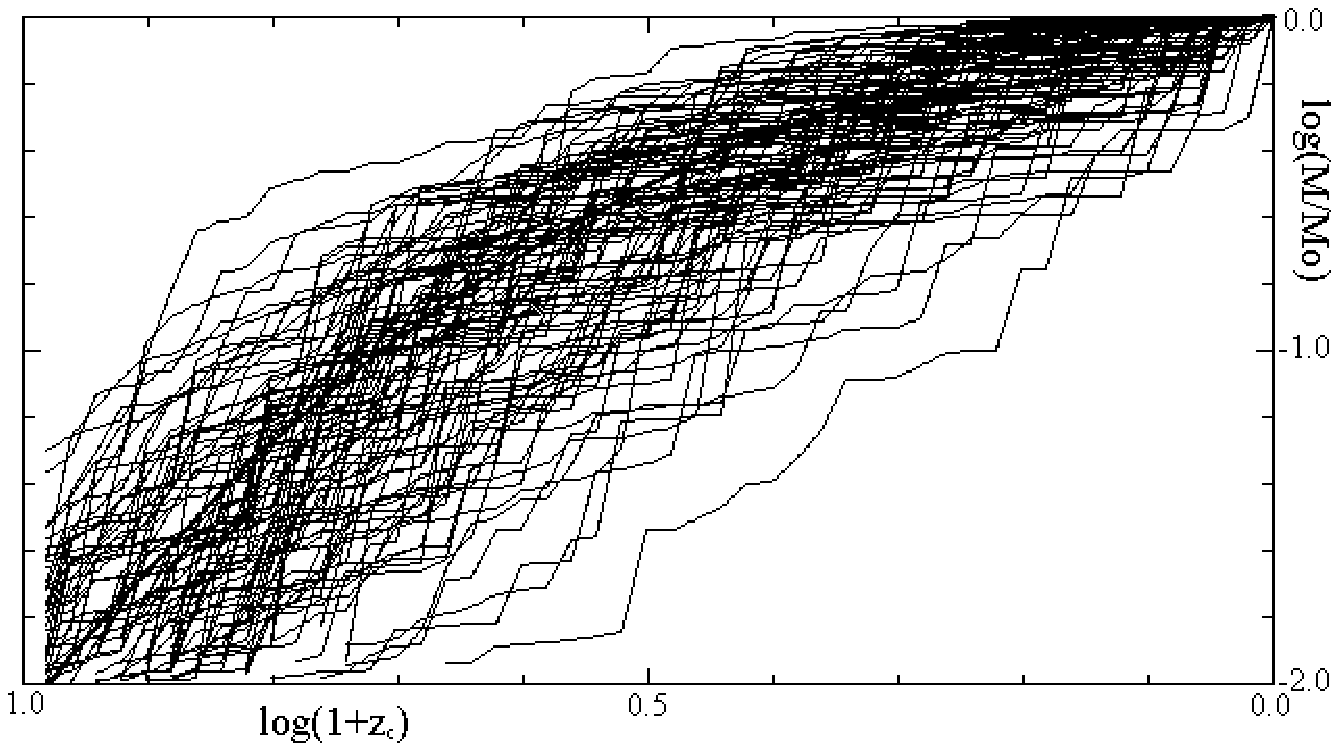}
\caption{The MAHs for a 5x$10^{11}M_{\odot }$ halo at z=0
($\Lambda CDM$) scaled to this mass. The average is given by the thick
line.}
\end{figure}
With a method based on the
conditional probability (Bower 1991, Lacey \& Cole 1993) we
follow the merging and mass aggregation history (MAH) of a given halo
in a linear regime. We are interested in haloes that
have not suffered during their entire evolution a major merger. Thus, we
exclude from our statistics the evolutionary tracks of those haloes that in
some time have collided with another halo with a mass greater than half of
its mass at that time (this reduce our sample of halo evolutionary
tracks roughly by
less than 20$\%$). Assuming spherical symmetry and using a statistical
approach to follow the shell-by-shell non-linear evolution of density 
fluctuations, we calculate the halo density profile evolution
(Avila-Reese, Firmani \& Hern\'andez 1998). The non radial
component of the kinetic energy is calibrated on the base of N-body
simulations. The angular momentum is calculated assuming a specific
value of the spin parameter $\lambda $, in agreement with the Zeldovich 
approximation. We have assumed an average value of $\lambda =0.05$
with a lognormal distribution where the $1-\sigma$
distribution is given at $\lambda =0.03$ and $\lambda =0.08$. Once a
mass shell is incorporated into the halo, its baryon fraction falls
onto the disk
assuming that the disk-halo feedback is negligible (we consider that
the SN energy feedback is localized into the disk in agreement with
the observational evidence). The gas shell is distributed across the
disk assuming rigid rotation of the shell and  detailed angular
momentum conservation during the gas collapse. The disk mass growth
produces a halo gravitational contraction that is calculated using an
adiabatic invariant technique.

\section{Results}
\begin{figure}
\vspace*{5.4cm}
\includegraphics{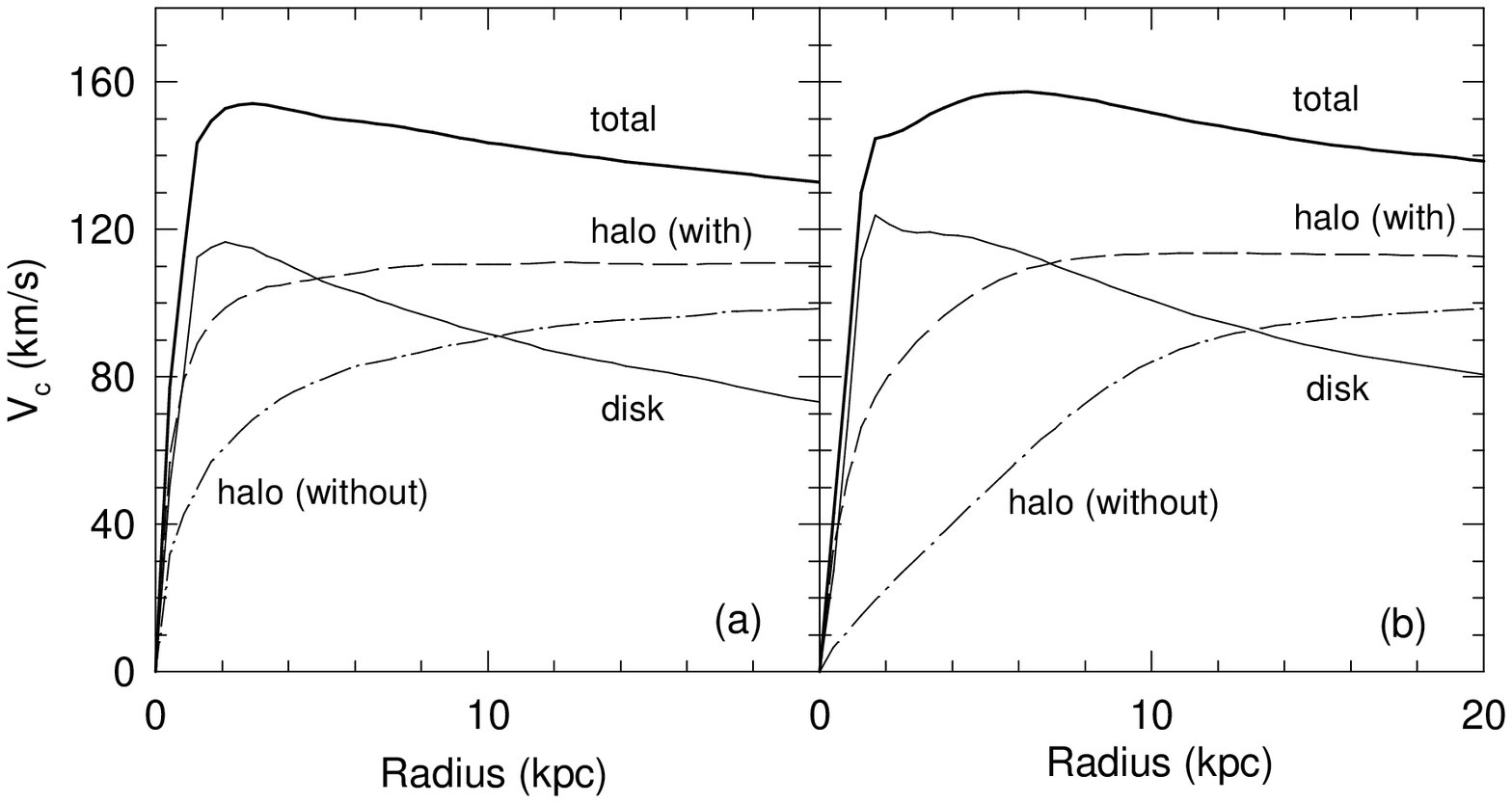}
\caption{Rotation curve decompositions. The DM halo contribution {\it
with} and {\it without} the gravitational contraction due to the disk
is shown. Panel (b) is for a DM halo with a near constant density core.}
\end{figure}
The MAHs for one halo of  5x$10^{11}M_{\odot }$ at $z=0$ $(\Lambda CDM)$ are
presented in fig. 1. We see here a wide range of MAHs which after
virialization produce a wide range of density profiles. The average
MAH is shown by the central thick line. This MAH produces the density profile 
obtained in N-body simulations by Navarro, Frenk and White
(1997). However, our method gives in a natural way the statistical
deviations from the average MAH, which lead to a rich variety
of halo profiles (Avila-Reese \etal 1998).  The rotation
curves obtained for the average MAHs, $\lambda =0.05$, and for the
masses 5x$10^{10}$, 5x$10^{11}$ and 5x$10^{12}M_{\odot} $ are
approximately flat. This explains the cosmological
nature of the conspiracy between baryon and dark matter in the flat profile
of the rotation curves. The decomposition of the rotation curve for the
model with 5x$10^{11}M_{\odot }$ ($\Lambda CDM$) is shown in fig. 2a, where
the halo gravitational contraction due to the disk is shown comparing the
halo rotation curve component {\it with} and {\it without} this
contraction. We note here a serious problem already pointed out in
Flores \& Primack (1994), Moore (1994), and Burkert (1995): 
the gravitational
contribution of the halo is dominant until the center. Even if the
uncertainty on the observational attempts to decompose the rotation curve is
large, we agree with Burkert's opinion that some physical process is
misunderstood here. Recent numerical results obtained by Kravtsov et
al. (1998) might be showing that the problem is not too serious,
and, comparing with our results, suggest that the merging process and
the slope of the power spectrum at the scale in consideration are
the responsible of giving rise to shallow cores in the DM halos. If we
assume that the central density halo profile before the gravitational 
contraction due to the disk agrees with the profiles observed in dwarf
and low surface brightness galaxies, then the final rotation curve
decomposition it looks like as it is shown in fig. 2b. 

\begin{figure}
\vspace*{5.3cm}
\includegraphics{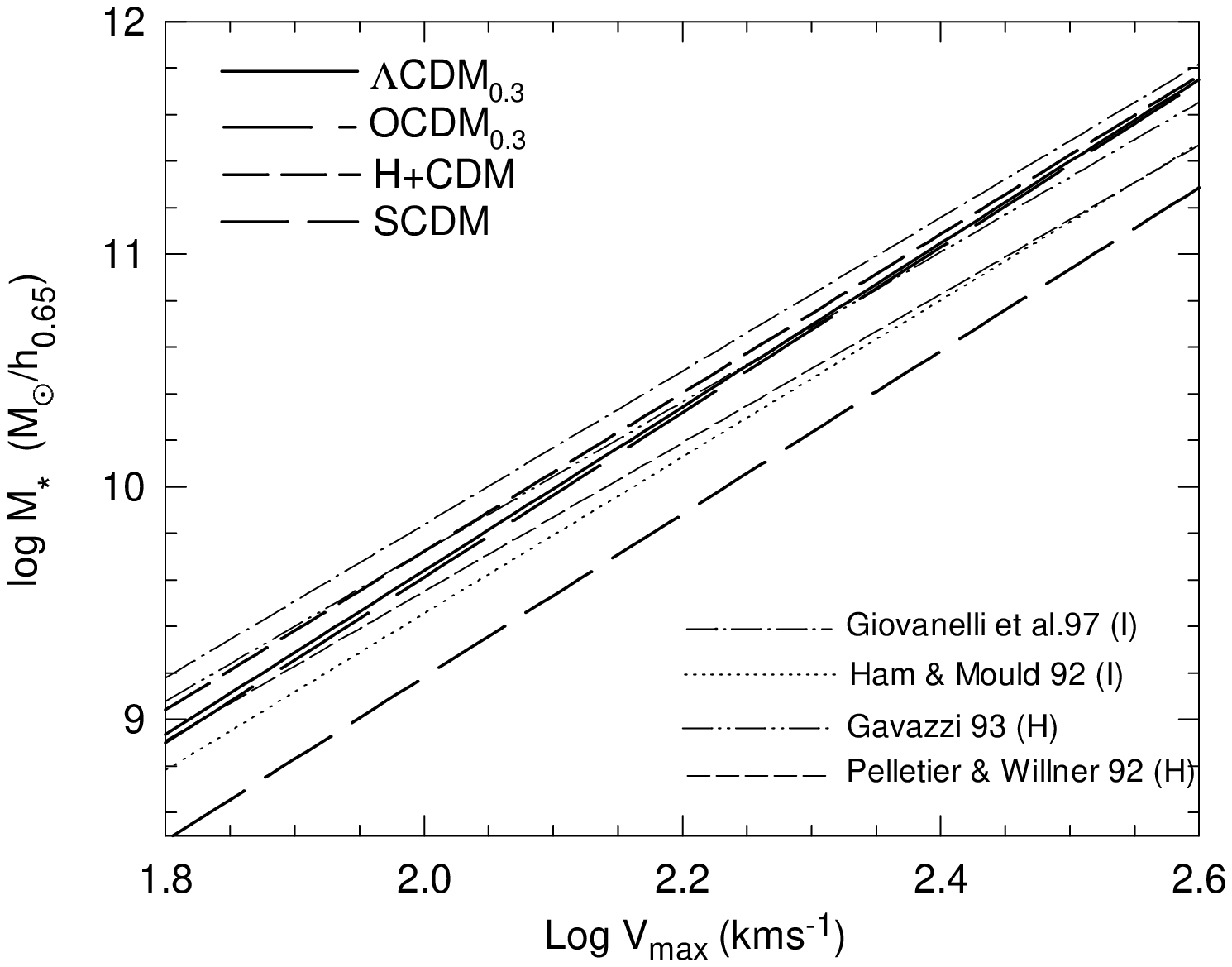}
\caption{The predicted TF relation for several cosmologies compared
with the observational data in the infrared. For the references see
Avila-Reese \etal 1998}
\end{figure}

The very simple mechanism we assume to
distribute the gas on the
disk, leads to a surface brightness distribution reasonably close to an
exponential profile. Observations show systematic blue radial
color gradients in spiral galaxies. Our models show a similar trend,
although some excessive gradient appears compared to the observations.
This problem may arise from the simple population synthesis technique that
works in our code or because a more complex process drives the gas infall
onto the disk. The final global properties of our models follow the
same trends of observed galaxies across the Hubble sequence: the
redder and higher the surface brightness are, the smaller is the gas fraction,
and the larger is the bulge-to-disk ratio. The three key factors that
determine the final properties of our models are the mass, the MAH and  
$\lambda$.
\begin{figure}
\vspace*{4.3cm}
\includegraphics{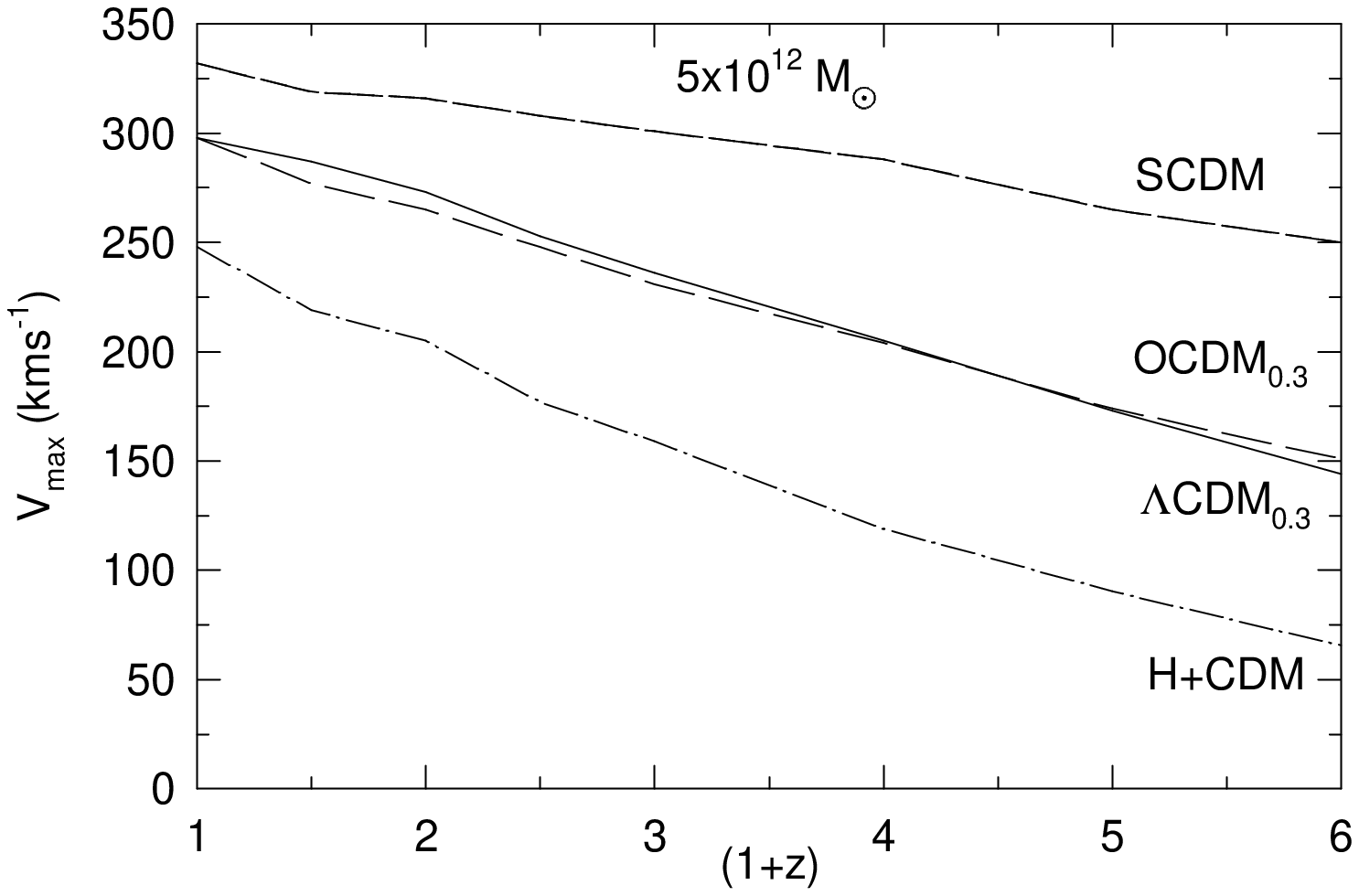}
\caption{The maximum rotation velocity {\it vs.} (1+z) for different 
cosmologies.}
\end{figure}

A main result of our work is the prediction of the TF
relation starting from the primordial conditions of the universe. The
density fluctuation power spectrum establishes the degree of the halo
concentration for each scale, and this leads to a relation between the
rotation velocity and the mass. Fig. 3 shows the TF relation predicted by
the different cosmologies, and the observational data obtained by different
authors in the infrared. The $SCDM$ model predicts rotation velocities that
exceed by a factor 1.4 the observations. This is due to the high power level
of the density fluctuations on galactic scale predicted by the $SCDM$
cosmology. A similar result has been obtained on larger scales by galaxy
counts. The $\Lambda CDM$, $OCDM$ and $H+CDM$ cosmologies predict TF
relations very close in slope and zero point with the observations. We have
explored the robustness of this result with respect to the baryon fraction
and the infall dissipative processes that retain the gas in the halo. In
both cases we find a surprising final robustness of our results, mainly due
to the gravitational contraction ``conspiracy'' of the DM
halo. Concerning the scatter around the TF relation, we have combined
the fluctuations due to the statistical nature of the MAHs and
$\lambda $ (considering the range of flat rotation curves), and we
have obtained for $\Lambda CDM$, $OCDM$ and $H+CDM$
an average scatter of roughly 0.4 mag that agrees
reasonably well with observations. For the $SCDM$ we predict an average
scatter of 0.6 mag which definitively exceeds the observed value. We
conclude that the TF relation represents an imprint of the primordial 
cosmological conditions on the galactic scales (see also Navarro et
al. 1997).
\begin{figure}
\vspace*{4.6cm}
\includegraphics{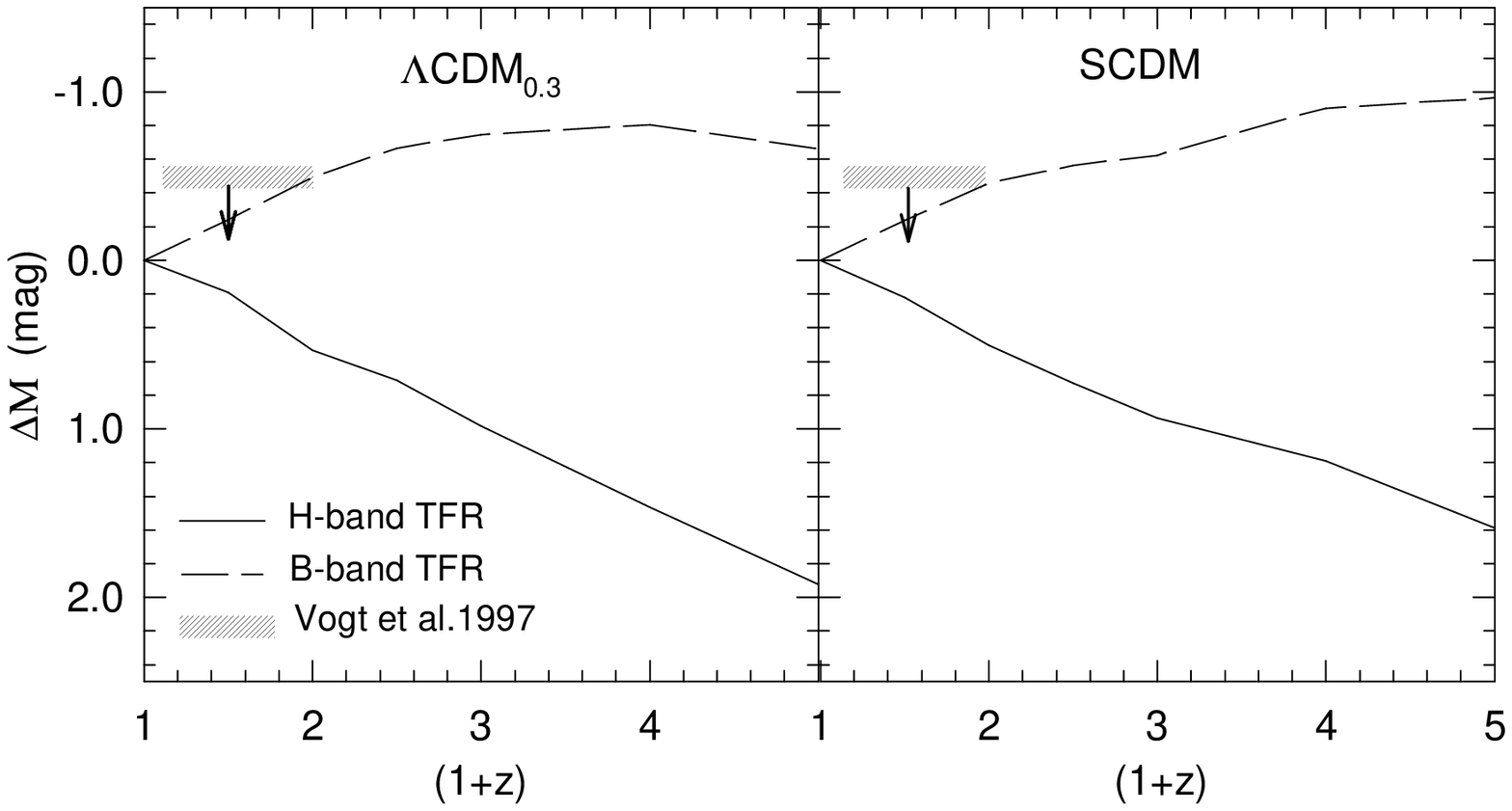}
\caption{The evolution of the TF zero-point in the H- and B-bands with
an upperlimit from the observations.}
\end{figure}

The \textit{semi-numerical} approach allows us to follow the evolution of an
individual galaxy, then we are able to predict how a galaxy appears at
different redshifts. Fig. 4 shows the behavior of $V_{\max }$ {\it
vs.} $z$. Here
we see the difficulty of $H+CDM$ to produce galaxies with high rotation
velocity in the redshift range proper of the damped $Ly\alpha $
absorbers (e.g.,
Klypin \etal 1995). In fig. 5 we show the evolution of the TF zero
point in the H and B bands, together with a B band upper limit derived
by Vogt \etal (1997) (the slope is almost constant). While in the
H-band the zero-point decreases with $z$, in the B-band the zero-point
remains almost constant because of the increase of the B-band
luminosity toward the past (Firmani \& Avila-Reese 1998). 
Finally fig. 6 displays the disk scale radius {\it vs.}(1+z). An
inside-out evolution is evident. It is important to remark that our
evolutionary tracks concern basically isolated galaxies, where the
environment may supply any amount of gas to a galaxy according to its
gravitational field. Our results are not representative of the average
conditions of the universe neither of more complex situations as the case of
galaxies in clusters. More work is planned in the future on this direction.

\section{Conclusions}
1) The semi-numerical models presented in this work support the viability of
an inside-out disk galaxy formation scenario, where the rate of gas
accretion on the disk is dictated mainly by the cosmological (hierarchical)
mass aggregation rate. 2) The TF relation is predicted as a
product of the cosmological initial conditions. The gravitational pull of
the luminous matter on the dark halo makes this relation robust with respect
to intermediate processes (cooling, feedback). 3) Concerning the predictive
abilities of different cosmologies with respect to galaxy formation and
evolution: i) $\Lambda CDM$ and $OCDM$  cosmologies are
able to predict many of the galaxy features up to intermediate redshift. ii) 
$SCDM$ is ruled out because is unable to predict the TF relation. iii) $H+CDM
$ is marginal because predicts a too late galaxy formation.

\begin{figure}
\vspace*{4.3cm}
\includegraphics{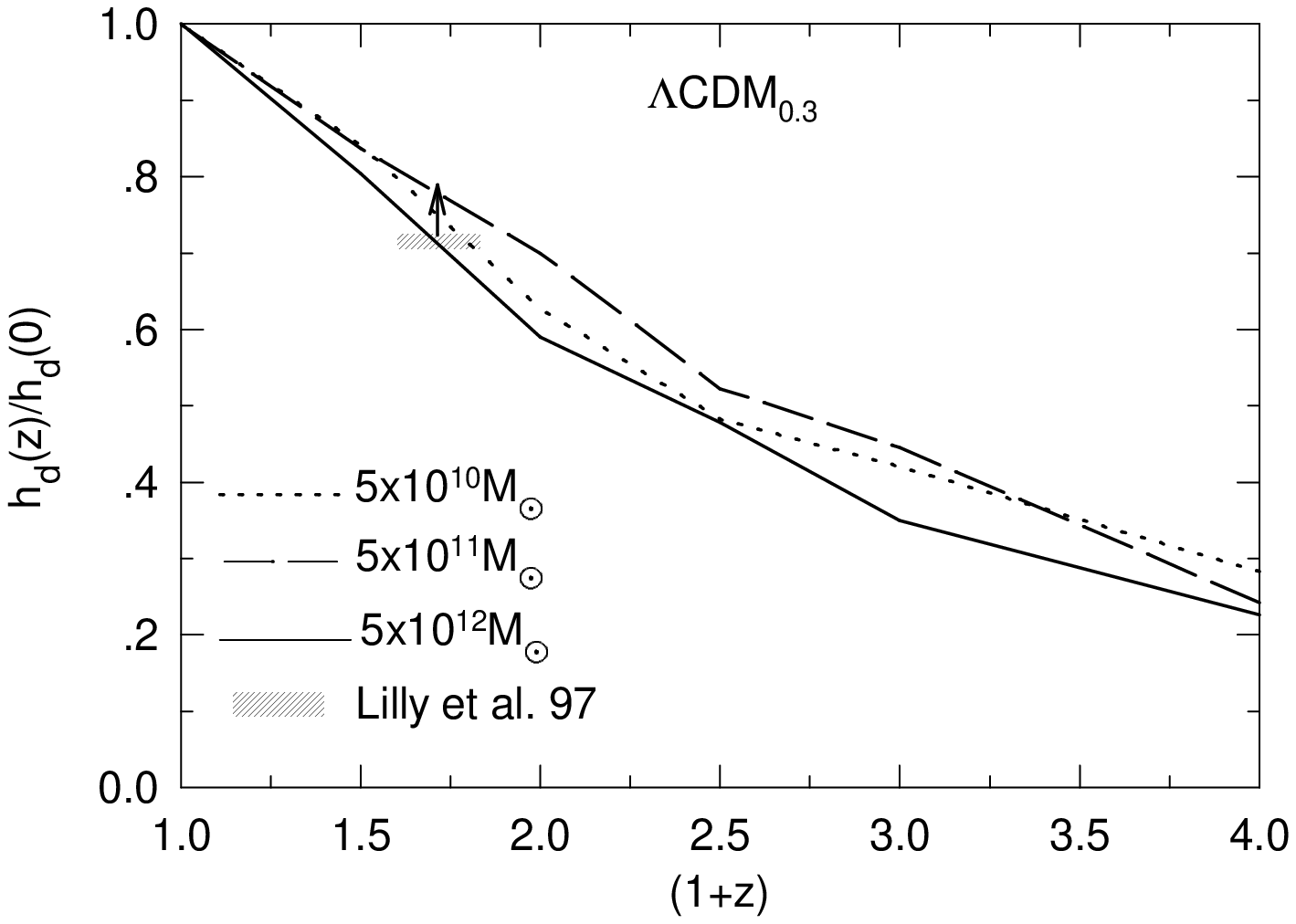}
\caption{The scale radius {\it vs.} (1+z) and a lower limit stablished by the
observations.}
\end{figure}

\end{document}